\documentclass[]{article}
\usepackage{graphics}
\newcommand{\ber}{\begin{eqnarray}}
\newcommand{\eer}{\end{eqnarray}}
\newcommand{\bea}{\begin{equation}}
\newcommand{\eea}{\end{equation}}
\def\be{\begin{eqnarray}}
\def\ee{\end{eqnarray}}
\def\bd{\begin{displaymath}}
\def\ed{\end{displaymath}}

\begin{document}

\title{
Many-particle many-hole states near the magic number 20: deformed and superdeformed states}

\author{Samit Bhowal\\ 
Department of Physics, Surendranath Evening College\\
24/2 Mahatma Gandhi Road, Kolkata- 700 009, India\\
Sankha Das\\
Department of Physics, MCKV Institute of Engineering\\
243 G.T. Road(N), Liluah, Howrah-711 204, India\\
G. Gangopadhyay\\
$^c$Department of Physics, University College of Science\\
 University of Calcutta,
92 Acharya Prafulla Chandra Road,\\
 Kolkata-700 009, India}
\maketitle

\abstract
{Many-particle many-hole states near the magic number 20 have been investigated
in the relativistic mean field formalism using the fixed configuration method.
Neutron particle-hole states in neutron-rich nuclei with $N\sim 20$, {\em i.e.} 
$^{30,32,34}$Mg,and $^{28,30}$Ne are studied to find out the ground state 
configuration. The ground state of $^{32}$Ne,Mg as well as possibly $^{34}$Mg 
comes out actually as particle-hole states. Proton-neutron excitation across the
shell gap at 20 in nuclei with $N=Z$, {\em i.e.} $^{32}$S, $^{36}$Ar,
$^{40}$Ca and $^{44}$Ti is investigated as possible origins of 
superdeformed configuration. Observed superdeformed bands in $^{36}$Ar 
and $^{40}$Ca can be described as many-particle many-hole states.}
\section{Introduction}

One of the most interesting topics in recent years in the study of low energy 
structure of light nuclei is the existence of highly deformed states in 
near-closed shell nuclei at relatively low excitation energy. Superdeformed (SD) 
bands have been observed in light $N=Z$ nuclei $^{36}$Ar\cite{Ar1,Ar2} and 
$^{40}$Ca\cite{Ca1,Ca2} as well as in $^{38}$Ar\cite{Ar38} with the band heads 
situated around 5 MeV excitation energy. Studies on $^{32}$Mg and the neighbouring 
neutron rich nuclei with $N\sim20$ showed very large B(E2) 
values for the $2^+_1\rightarrow 0^+_1$ transition indicating a large 
quadrupole deformation \cite{Mg32,Mg321,Mg322,Mg34}. The existence of highly 
deformed solutions near the magic number 20 is now understood to be a result of 
many-particle many-hole excitation across the shell gap. 
From the spherical shell model perspective, particles are promoted from 
the $sd$ shell to the f$_{7/2}$ orbital across the shell gap at $N,Z = 20$. 
When the prolate deformation is 
large, the oblate deformation driving last orbitals of the $sd$ shell are higher in 
energy than the low $\Omega$ orbitals coming from the $f_{7/2}$ orbital. 
Thus a state may be formed with a large prolate deformation at a relatively 
low excitation energy by filling up the single particle levels from below. 

In neutron rich nuclei around $^{32}$Mg, a pair of neutrons may be excited across the 
$N=20$ shell gap resulting in the $2\hbar\omega$ configurations. 
In $Z=10-12$ and $N=19-22$ nuclei, recent shell model calculations 
\cite{SM1,SM2,SM3,SM4}  predict these
$2\hbar\omega$ states to lie energetically below the normal $0\hbar\omega$
states. In the Nilsson model picture, the shell gap 
vanishes at large deformation for neutrons. This so called `island of inversion'
where the $2\hbar\omega$ states lie below the normal states have been 
extensively investigated both theoretically and experimentally. Different 
mean field calculations \cite{hfb1,hfb2,Lal,PM} differ in their predictions
for the ground state configurations of different nuclei in this region. There is 
considerable disagreement among the different calculations regarding the boundary of 
the island of inversion. Another area where the calculations do not agree is whether
the neutron and proton distributions are identical in these nuclei. Pritychenko
{\em et al.} have measured the transition probabilities $B(E2;0_{gs}^+\rightarrow 2^+_1)$
in the nuclei $^{26,28}$Ne and $^{30,32,34}$Mg to study the role of these intruder 
configurations in an effort to determine the boundary. They conclude that the
ground state of the nuclei $^{26}$Ne and $^{30}$Mg are well understood in terms of normal
$0\hbar\omega$ configurations. On the other hand Chist\'{e} {\em et al.}
have performed inelastic scattering experiment to measure the charge and mass 
deformation of $^{30,32}$Mg. They conclude that both the isotopes show similar type of 
structure involving $2\hbar\omega$ configurations and there is no significant difference 
between the proton and the neutron distributions.

The SD band \cite{Ca1,Ca2} in $^{40}$Ca is based on the $ J^\pi=0^+$ state at 5.21 MeV. 
In $^{40}$Ca, low lying states with large collectivity have been explained in terms of
neutron-proton excitation across the shell gap at $N,Z=20$. Fortune {\em et al.}
\cite{F1,F2} have suggested the dominant configurations of the $J^\pi=0^+$ states at 0 
MeV, 3.35 MeV and 5.21 MeV to be $0p-0h$, $4p-4h$ and $8p-8h$, respectively. A full
shell model calculation involving both the $sd$ shell and the $fp$ shell is still beyond
the present day capabilities. A recent truncated shell model calculation \cite{Ca} has
confirmed that the SD band in $^{40}$Ca is indeed based on $8p-8h$ configuration. Cranked 
Hartree-Fock-BCS method has been applied by Bender {\em et al.} to study the
ND and SD states in $^{32}$S, $^{36,38}$Ar and $^{40}$Ca\cite{chf}.  
In Ref. \cite{Ca1} cranked relativistic mean field  theory without pairing was applied 
to study the 
highly deformed states in $^{40}$Ca.  Zheng {\em et al.}\cite{Z1,Z2} have performed fixed 
configuration deformed Hartree-Fock calculation to study the deformed states in a number 
of nuclei including $^{40}$Ca and $^{44}$Ti. The origin of the SD band is predicted to be an 
$8p-8h$ configuration in all the theoretical investigations. Configuration dependent cranked 
Nilsson-Strutinsky and truncated shell model calculations were performed \cite{Ar1} in
highly deformed states in $^{36}$Ar. All calculations agree that in $^{36}$Ar, these states are 
expected to be based on $4p-8h$ configuration coming from the excitation of two protons and 
two neutrons across the shell gap.

The aim of the present work is to study these large deformation 
multiparticle-hole states. We have applied the fixed configuration deformed 
mean field approach, a method applied earlier in this mass region by Zheng {\em et al.}
\cite{Z1,Z2} using a Skyrme Hartree Fock model. We propose to use the same
method using deformed Relativistic Mean Field (RMF) approach. This approach has been very
successful in describing various features of nuclear structure. Ground state 
properties like binding energy, deformation, magnetic moment, isotopic shift and 
nuclear radius has been calculated with considerable accuracy\cite{RMF}. It has 
also succeeded in describing excited states like normal deformed (ND) and 
SD bands, giant dipole resonances, etc. One of the prominent 
successes of RMF is the reproduction of the spin-orbit splitting naturally. 
It has been observed in \cite{Z1} that  the spin-orbit interaction plays a major
role in the calculation for the excited states. The RMF approach naturally 
accommodates a correct spin-orbit interaction and thus, is expected to give a good 
description of the multiparticle-hole states. We have not come 
across any RMF calculation using the fixed configuration approach.

\section{Calculation}

Relativistic Mean Field theory is well known and has been described in
detail elsewhere\cite{RMF}. We have employed the force NL3\cite{Ring2}
which was obtained by fitting the binding energy, charge radii and neutron radii of
a number of spherical singly and doubly closed nuclei. This was explicitly designed for
treatment of variation in isospin and has proved to be very useful in describing the 
ground state properties throughout the periodic table. Some of the nuclei studied in
the present work, {\em{e.g.}} $^{32,34}$Mg and $^{30}$Ne are away from the stability 
valley and should be well described using the force NL3.  We assume axial symmetry and 
reflection symmetry and work using deformed harmonic oscillator basis.  
The method of calculation described in Refs. \cite{Gam,Gam1} has been
followed. 
Throughout this work, 12 Fermion and Boson shells have been used for the calculation. 
The quadrupole charge deformation parameter $\beta_C$ is obtained from
the intrinsic charge quadrupole moment $Q_0$ using the prescription
\be Q_0=\sqrt{\frac{16\pi}{5}} \frac{3}{4\pi}ZR_0^2\beta_C\ee
where $R_0=1.2A^{1/3}$, $A$ being the mass number. The mass deformation 
parameter $\beta_A$ can be similarly defined.

The fixed configuration calculation is essentially very simple. The 
states studied in the present calculation are particle-hole states only in the
sense of the spherical shell model. As already explained, at large deformation, the
levels coming from the higher $fp$ shell cross the upper levels of the $sd$ shell 
giving rise to the configuration of these states. The level 
sequence  obtained from Nilsson-like diagrams are used to construct solutions
with large deformation. For example, in Fig. 1, the single particle neutron 
orbitals in $^{40}$Ca are plotted as a function of the quadrupole 
deformation parameter $\beta_C$. Single particle proton orbitals also show a 
similar kind of 
structure. The orbits are identified by the asymptotic quantum number 
$\Omega^\pi[Nn_3\Lambda]$ at large deformation and by $L_j$ at zero deformation.
For N=20, there are two crossings. The first  one is around $\beta_C=0.3$ 
between the $3/2^+[202]$ orbit and the $1/2^-[330]$ orbit. The second crossing 
is at $\beta_C\sim 0.6$ between the $1/2^+[200]$ orbit and $3/2^-[321]$ orbit. In the 
fixed configuration approach, to obtain the $4p-4h$ solution, the $3/2^+[202]$ 
orbits are kept empty and the $1/2^-[330]$ orbits are kept filled for both types 
of nucleons. The relativistic Hartree equations are then solved iteratively to obtain
the binding energy and the deformation self-consistently. The resulting solution is 
expected to have a deformation beyond 0.3. Similarly, the $8p-8h$ solution obtained by 
further keeping the $1/2^+[200]$ orbits empty and the $3/2^-[321]$ orbits filled are 
expected to have a deformation beyond 0.6. The Hartree-BCS equations are then 
self-consistently solved to obtain the lowest energy solution. The solutions obtained 
in this way are of course intrinsic solutions, i.e. they contain contributions from all 
possible $J$ values in the band. We are principally interested in the $J^\pi=0^+$ state,
i.e. the band head of the K=0 bands. To obtain the energy of the $0^+$ state
we have used the cranking model formula
\be {\rm E}(0^+)={\rm E}_{in}-\frac{<J^2_\perp>}{2I_{cr}}\ee
where the moment of inertia $I_{cr}$ is calculated using the cranking model formula
\cite{MI} and $J^2_\perp=J^2_x+J^2_y$. Here, E$_{in}$ refers to the energy of the 
intrinsic solution. This is a very important correction as
at a large deformation, it can be quite large while being zero for
spherical configurations. For example, angular momentum projection in HF calculation 
in the superdeformed minimum of $^{40}$Ca lowers the bandhead by 4 MeV\cite{chf}.

Pairing has been included in our calculation in the form of BCS method in
the constant gap approach. We have taken both the proton and the neutron gaps to be
equal to 1.0 MeV except for the closed shell configurations. In the light 
nuclei studied, the pairing energy is quite small. We have verified 
that our essential conclusions remain unaltered for reasonable variation of 
the gap parameters.

\section{Results}

We first present our results for neutron excitation across the $N=20$ shell gap in  a 
number of neutron rich nuclei near $^{32}$Mg.  Next we discuss the proton-neutron 
excitation across the shell gap $N=Z=20$ in the $N=Z$ nuclei,  $^{32}$S, 
$^{36}$Ar, $^{40}$Ca and $^{44}$Ti. We present the results of our calculation in 
Table I. Here E(0$^+$) refers to the energy of the $0^+$ state belonging
to the particular configuration and $\beta_C$ and $\beta_A$ refer to charge and
mass quadrupole deformation parameters, respectively.

The nucleus $^{32}$Mg has $N=20$ and $Z=12$. In general, a ground state 
calculation without particle-hole excitation predicts the ground state 
to be spherical. This has been observed in earlier relativistic calculations 
also \cite{Lal,PM}. In particular, although the deformation of a 
number of light nuclei was calculated\cite{PM,PM1} with some degree of accuracy in the
relativistic hybrid derivative coupling model, the 
ground state of $^{32}$Mg is still found to be spherical in this model. 
In the present
calculation, we find that the spherical minimum has a binding energy
250.97 MeV. Next we study the $2p-2h$ neutron excitation across the magic number
20. This corresponds to two holes in the $\nu3/2^+[202]$ state and two particles 
in the $\nu1/2^-[330]$ state. The intrinsic solution has the binding energy 248.93 MeV. 
After the correction due to angular momentum, the binding energy of the deformed 
$2p2h$ 0$^+$ 
state comes out to be 252.50 MeV. Thus the $\nu^2-\nu^{-2}$ state becomes the actual 
ground state. The experimentally measured binding energy is 249.69 MeV \cite{WA}. 
The charge and the mass quadrupole deformation parameters 
 come out to be 0.457 and 0.492, respectively. The different experimental 
measurements on charge deformation parameters give the observed values
$\beta_C=0.512\pm 0.044$ \cite{Mg32} in intermediate energy Coulomb excitation and 
$\beta_C=0.61\pm0.04$ from inelastic scattering experiment
\cite{Mg322}. From the results of another Coulomb excitation study \cite{Mg321}, 
a slightly lower value of $\beta_C=0.438\pm 0.046$ can be derived. Thus the 
experimentally measured values are reasonably close to the theoretical result. The 
proton and the neutron distributions come out to be similar. This is in agreement 
with the measurements of Chist\'{e} {et al.} \cite{Mg322} who have found no hint 
of any large decoupling between the proton and the neutron 
distribution. Our result also agrees with shell model calculations \cite{SM2,SM4} as
well as HFB results \cite{hfb2}. In contrast, another HFB calculation \cite{hfb1}
predict the ground state to be spherical, i.e. of normal configuration.
We have also examined the $\nu^4-\nu^{-4}$ excitation but could not
obtain any converged self-consistent solution.

We have studied a few nearby even-even nuclei in the present formalism to
find out the boundary of the island of inversion. In $^{30}$Mg, the 
ground state turns out to be a 0$\hbar\omega$ state with no $p-h$ excitation.
The binding energy of the ground state is calculated to be 243.66 MeV.
The experimentally measured binding energy is 241.63 MeV\cite{WA}.
The $\nu^2-\nu^{-4}$ state is obtained by exciting two neutrons from the
$\nu1/2^+[200]$ orbital to the $\nu1/2^-[330]$ orbital across the $N=20$ gap.
The binding energy of the $2p4h$ 0$^+$ state comes out to be 242.17 MeV, approximately 
1.5 MeV above the ground state. The calculated charge and mass deformation 
values are 0.522 and 0.601, respectively. Inelastic scattering gives the
experimental value $\beta_C=0.52\pm 0.04$ \cite{Mg322}. Coulomb excitation 
experiment gives a somewhat lower value $\beta_C=0.43\pm 0.19$\cite{Mg321}.
As already mentioned, Pritychenko {\em et al.}\cite{Mg321} have suggested that 
the ground state of $^{30}$Mg to be a $0\hbar\omega$ state while Chist\'{e}
{\em et al.}\cite{Mg322} have concluded it to be a $2\hbar\omega$ state.
The energy systematics observed in our calculation seem to support the
former alternative although a configuration mixing calculation is required
to reach a definite conclusion.

In $^{34}$Mg, the two solutions corresponding to the $0\hbar\omega$
and $2\hbar\omega$ configurations come out to be very close in energy.
The latter configurations has two holes in the $\nu3/2^+[202]$ orbital and two 
particles in the $\nu1/2^-[330]$ and $\nu3/2^-[321]$ orbitals each.
The $0\hbar\omega$ state has binding energy 260.48 MeV and the charge (mass)
deformation values 0.345 (0.327). The binding energy for the $2\hbar\omega$ 
state is 260.42 MeV and the $\beta_C$ and $\beta_A$ values are 0.511
and 0.562, respectively. Coulomb excitation experiment provides a upper
bound for the charge deformation $\beta_C<0.599$ \cite{Mg321}. The experimental 
ground state binding energy is 256.59 MeV
\cite{WA}. A more accurate treatment, particularly of the pairing correlation,
may show that the $2\hbar\omega$ configuration is actually the ground state.

In $^{30}$Ne, the $0\hbar\omega$ solution turns out to be spherical with the 
binding energy 215.93 MeV. The $2\hbar\omega$ 0$^+$ state has the binding energy 
216.62 MeV, making it the ground state. The experimental binding energy
is 212.08 MeV.
In $^{28}$Ne, the situation is similar to  $^{30}$Mg. The $2\hbar\omega$ state 
turns out to be 1.75 MeV above the
$0\hbar\omega$ state. The calculated binding energy of the ground state
is 210.57 MeV while the measured value is 208.82 MeV. The neutron configurations 
of the $2\hbar\omega$ states
in $^{30}$Ne and $^{28}$Ne are identical with $^{32}$Mg and $^{30}$Mg,
respectively. We have also studied $^{32}$Ne but could not obtain any 
self-consistent solution corresponding to the $\nu^4-\nu^{-2}$ state.
We have also looked for low-lying $2\hbar\omega$ states in $^{34}$Si
but could not find any self-consistent solution.

Among the $N=Z$ nuclei, we start with  the doubly magic nucleus $^{40}$Ca.
The binding energy of the ground state is calculated to be 341.92 MeV, in good
agreement with the experimental value 342.05 MeV. The ground state is spherical.
A number of $np-nh$ states has been studied in \cite{Z1} under the SkHF formalism.
They observed that for $n$=2 to 8, although $\beta$ increases linearly with $n$,
the solutions are nearly degenerate in energy. The $8p-8h$ state has been 
identified with the SD band whose band head lies at 5.21 MeV excitation energy.
The $4p-4h$ state has been identified with the ND band whose band head lies at
3.35 excitation energy. However for some of the Skyrme parameterizations, the
$0^+$ state of the ND band was found out to lie higher than that of the SD band
in contradiction with experimental results. 

In the present calculation, we have 
concentrated on the $4p-4h$ excitation and the $8p-8h$ excitation. The different 
configurations have been discussed earlier. In our 
calculation also, the ND band comes up very high in energy. In contrast, the 
excitation energy of the $0^+$ state of the $8p-8h$ band comes out to be 5.86 MeV.
The 
transition quadrupole moment values also show a similar trend. The ND band
quadrupole moment is calculated to be 1.14 eb, large compared to the 
experimental value $Q_0=0.74\pm0.14$eb\cite{Ca1}. However, the SD band quadrupole moment
comes out to be 1.89 eb, in excellent agreement with the experimental value
$Q_0=1.80^{+0.39}_{-0.29}$eb\cite{Ca1}. In contrast, the cranked RMF
calculation without pairing \cite{Ca1} predicts $Q_0$ to be 2.0 eb.

One of the problems in getting a good description of the ND band may be the near 
degeneracy of the $np-nh$ states for even $n$. This degeneracy was observed in the present
calculation also.  A proper calculation should take into consideration the mixing
between all the $0^+$ states coming from the different $np-nh$ configurations. This 
may provide a more accurate description.

The ground state of the nucleus $^{36}$Ar turns out actually to be oblate, This is in 
agreement with the number and angular momentum projected HF calculation of Bender 
{\em et al.}\cite{chf}.
The calculated binding energy is 306.36 MeV. In comparison, the experimental 
value is 306.72 MeV. For the $4p-8h$ state, we remove two protons and two neutrons
from the $3/2^+[202]$ orbit and put them in the $1/2^-[330]$ orbit. The  energy of the 
excited $0^+$ state after correction comes out to be 300.39 MeV. The excitation energy 
of the band head is thus 5.97 MeV. In comparison, this energy is calculated to be 5.90 
MeV in Ref. \cite{chf}.  The experimental value of the band head is
at 4.33 MeV. The experimental transition quadrupole moment is $1.18\pm 0.09$ eb \cite{Ar2}. 
The calculated intrinsic quadrupole moment is 1.37 eb.

We also study two other nuclei with $N=Z$, $^{32}$S and $^{44}$Ti where the SD band 
has not been observed. In $^{32}$S, the experimental binding energy of the
ground state is 271.78 MeV. We tried to look for solutions with 12 holes in
the $3/2^+[202]$, $1/2^+[200]$ and $5/2^+[202]$ orbitals and  4 particles in
the $1/2^-[330]$ orbitals.  The excitation energy of the solution comes out to be 6.22 MeV. 
and the deformation  to be very high, $\beta\sim 1.0$. We could not get any 
converged self consistent solution for only proton or neutron excitation.
In $^{44}$Ti, the experimental binding energy is 375.47 MeV while the
calculated value comes out to be 372.48 MeV. We have obtained a 
$8p-4h$ solution with the $3/2^+[202]$ orbitals empty and the $1/2^-[330]$ and the 
$3/2^-[321]$ orbitals filled. The excitation energy of the corresponding band
comes out to be 3.36 MeV and the charge deformation 0.530.
In Ref. \cite{Z1}, the excitation energy of the $8p-4h$ band is found to be
5.6 MeV and the deformation value $\beta=0.41$. 

\section{Conclusions}
We have observed that the nuclei $^{32}$Mg, $^{30}$Ne, and possibly $^{34}$Mg
belong to the island of inversion. As for the nuclei $^{30}$Mg and $^{28}$Ne,
both nuclei with $N=18$, the ground states are predicted to be moderately deformed
$0\hbar\omega$ states while the $2\hbar\omega$ states have very large deformation.
Measurements indicate that the ground state in these two nuclei are actually
strongly deformed. The present calculation is unable to describe the ground state 
properties of these nuclei. In all the strongly deformed systems, we find that
the proton and the neutron distributions are not decoupled, in agreement with 
Chist\'{e} {et al.} \cite{Mg322}.

In $^{40}$Ca and $^{36}$Ar, the SD band seems to be reasonably described in the
present formalism. Particularly, the prediction of the intrinsic quadrupole
moments is in excellent agreement with experimental measurements. However, in 
$^{40}$Ca, the normal deformed configuration comes too high in energy. 
A source of error in the excitation energy of the bandhead may be the neglect of 
configuration mixing in the particle-hole bands. 
Surprisingly, we find no indication of any SD minimum in $^{32}$S. 

We have included pairing through BCS approximation. The nuclei in the island of 
inversion have very large neutron excess and are close to the drip line. A more 
accurate treatment should involve the Bogoliubov approach and possible
coupling to the continuum as well as particle number and angular momentum projection.

The calculations were done using the computer facilities provided under the
DSA Programme by the University Grants Commission, New Delhi.

\clearpage
\begin{table*}[h]
\caption{Theoretical results obtained in the present calculation. 
See text for details. Configurations denoted normal are $0\hbar\omega$
solutions.}
\center
\begin{tabular}{cccrcccc}\hline
Nucleus & Configuration & E$_{in}$& $J_\perp^2/\hbar^2$&$I_{cr}/\hbar^2$&E.(0$^+)$
&$\beta_C$&$\beta_A$\\
&&MeV&&MeV$^{-1}$&MeV&\\\hline
$^{32}$Mg & normal & -250.97& &&-250.97&0.000& 0.000\\
          & $\nu^2-\nu^{-2}$ & -248.93&33.67&4.73&-252.50& 0.457 & 0.492\\\hline
$^{30}$Mg &normal  &  -240.68 & 10.95&1.84&-243.66&0.285 & 0.230\\
          &$\nu^2-\nu^{-4}$  &  -237.75 &36.93&4.18&-242.17&0.522&0.601\\\hline
$^{34}$Mg & normal & -257.62 & 21.53 & 3.77 & -260.48& 0.345 & 0.327\\
          &$\nu^4-\nu^{-2}$  & -255.81 & 45.20&4.91&-260.42&0.511& 0.562\\\hline
$^{30}$Ne & normal & -215.93 & &&-215.93&0.000 & 0.000 \\
          &  $\nu^2-\nu^{-2}$ & -213.06 &27.62&3.88& -216.62&0.444&0.502\\\hline
$^{28}$Ne &normal  & -208.61 &5.32&1.35&-210.57&0.202 & 0.162\\
          &$\nu^2-\nu^{-4}$  & -204.72 & 29.35&3.58&-208.82&0.499&0.619\\\hline\hline
$^{40}$Ca & normal & -341.92 &&&-341.92& 0.000 & 0.000 \\
          & $\nu^2\pi^2-\nu^{-2}\pi^{-2}$ & -328.43&42.70&5.50&-332.31 & 0.449 & 0.442 \\
          &  $\nu^4\pi^4-\nu^{-4}\pi^{-4}$ & -329.80&70.29&5.61&-336.06 & 0.743 & 0.733 \\\hline
$^{36}$Ar &normal  & -302.59 & 12.20 & 1.62& -306.36 & -0.204 & -0.201\\
          & $\nu^2\pi^2-\nu^{-4}\pi^{-4}$  &-295.25&47.65&4.64&-300.39& 0.642 & 0.632\\\hline
$^{32}$S  & normal & -266.12 &10.19&1.43&-269.68&0.240 & 0.236 \\
          & $\nu^2\pi^2-\nu^{-6}\pi^{-6}$ & -256.45&64.52&4.60&-263.46 & 1.030& 1.017 \\\hline
$^{44}$Ti & normal & -370.80&11.03&3.28&-372.48&0.135&0.133\\
          & $\nu^4\pi^4-\nu^{-2}\pi^{-2}$ & -364.45&60.40&6.46&369.12&0.530&0.523 \\\hline
\end{tabular}
\end{table*}  
\clearpage 
\centerline{\bf List of Figure captions}
\vskip 2cm
\begin{figure}[h]
\vskip 0.5cm
\caption{\label{fig2}The neutron single particle energy levels
in $^{40}$Ca as a function of quadrupole deformation $\beta_C$.
The levels are indicated by the spherical quantum numbers at 
zero deformation and the asymptotic quantum numbers at large deformation.
Continuous (dashed) lines represent positive (negative) parity 
levels.}
\end{figure}


\begin{thebibliography}{99}

\bibitem{Ar1}C.E. Svensson {\em et al}, Phys. Rev. Lett {\bf 85}, 2693 (2000).
\bibitem{Ar2}C.E. Svensson {\em et al.}, Phys. Rev. C {\bf 63}, 061301(R) (2001).
\bibitem{Ca1} E. Ideguchi {\em et al.}, Phys. Rev. Lett. {\bf 87}, 222501 (2001).
\bibitem{Ca2} C.J. Chiara {\em et al.}, Phys. Rev. C {\bf 67}, 041303(R) (2003).
\bibitem{Ar38}D. Rudolph {\em et al.}, Phys. Rev. C {\bf 65}, 034305 (2002).
\bibitem{Mg32}T. Motobyashi {\em et al.}, Phys. Let. {\bf 346B}, 9(1995).
\bibitem{Mg321}B.V. Pritychenko {\em et al.}, Phys. Let. {\bf 461B} 322 (1999).
\bibitem{Mg322} V. Chist\'{e} {\em et al.}, Nucl. Phys. {\bf A682} 161c (2001).
\bibitem{Mg34} H. Iwasaki {\em et al.}, Phys. Lett. {\bf B 522}, 227 (2001).
\bibitem{SM1} E.K. Warburton, J.A. Becker and B.A. Brown, Phys. Rev. C{\bf 41}
1147 (1990).
\bibitem{SM2} N. Fukunishi, T. Otsuka and T. Sebe, Phys. Lett. {\bf B296}, 279 (1992).
\bibitem{SM3} A. Poves, J. Retamosa, Nucl. Phys. {\bf A 571}, 221 (1994).
\bibitem{SM4} E. Caurier, F. Nowacki, A. Poves and J. Retamosa , Phys. Rev. C {\bf 58}, 
2033 (1998).
\bibitem{hfb1}J. Terasaki, H. Flocard, P.-H Heenen and P. Bonche, Nucl. Phys. 
{\bf A 621}, 706 (1997).
\bibitem{hfb2} R. Rodriguez-Guzman, J.L. Egido and L.M. Robeldo, Phys. Lett {\bf B474}, 15 (2000).
\bibitem {Lal} G. A. Lalazissis, A. R. Farhan and M. M. Sharma,
Nucl. Phys. {\bf A628}, 221 (1998).
\bibitem{PM} P. Mitra, G. Gangopadhyay and B. Malakar, Phys. Rev. C
{\bf 65}, 034329 (2002).
\bibitem{F1} H.T. Fortune, R.R. Betts, J.N. Bishop, M.N.I. Al-Jadir and R. Middleton,
Phys. Lett. {\bf 55B}, 439 (1975).
\bibitem{F2} H.T. Fortune, M.N.I. Al-Jadir, R.R. Betts, J.N. Bishop and R. Middleton,
Phys. Rev. C {\bf 19}, 756 (1979).
\bibitem{Ca} E. Caurier, F. Nowacki, A. Poves and A. Zuker, 
{\em The superdeformed
excited band of $^{40}$Ca}, preprint nucl-th/0205036 (2002).
\bibitem{chf} M. Bender, H. Flocard and P.-H. Heenen, Phys. Rev. C{\bf 68},  
044321 (2003). 
\bibitem{Z1} D.C. Zheng, L. Berdichevsky and L. Zamick, Phys. Rev. C {\bf 38},
432 (1988).
\bibitem{Z2} D.C. Zheng, L. Zamick and L. Berdichevsky, Phys. Rev. C {\bf 42},
1004 (1990).
\bibitem{RMF} See {\em e.g.} P. Ring, Prog. Part. Nucl. Phys., {\bf 37},
13 (1996).
\bibitem{Ring2} G.A. Lalazissis, J. K\"{o}nig and P. Ring, Phys. Rev C 
{\bf 55}, 540 (1997).
\bibitem{Gam} Y. K. Gambhir, P. Ring and A. Thimet, Ann. Phy. (N.Y.)
{\bf 198}, 132 (1990).
\bibitem{Gam1}P. Ring, Y.K. Gambhir and G.A. Lalazissis, Comput. Phys. Commun.
{\bf 105}, 77 (1997).
\bibitem{MI} S.G. Nilsson and O. Prior, Mat.-Fys. Medd. Dan Vidensk. Selsk. 
{\bf 32}, No. 16 (1960).
\bibitem{PM1} P. Mitra, B. Malakar and G. Gangopadhyay,
Int. Jour. Mod. Phys. E {\bf 10}, 475 (2001).
\bibitem{WA} G.A. Audi and A.H. Wapstra, Nucl. Phys. {\bf A595}, 409 (1995).

\end{thebibliography}
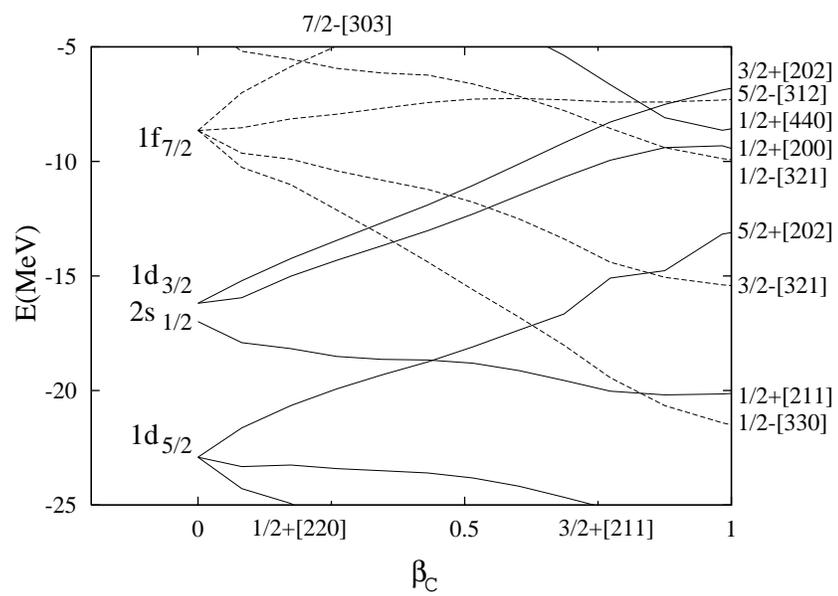
\end{document}